\documentclass[10pt]{emulateapj}
\usepackage{pslatex}
\usepackage{graphicx}
\usepackage{natbib}
\begin{document}

\title{Alfv\'en Wave Turbulence and Perpendicular Ion Temperatures in Coronal Holes}

\author{Benjamin D. G. Chandran}


\affil{Space Science Center and Department of Physics, University of New Hampshire, Durham, NH 03824; benjamin.chandran@unh.edu}

\begin{abstract}
  Low-frequency Alfv\'en-wave turbulence causes ion trajectories to
  become chaotic, or ``stochastic,'' when the turbulence amplitude is
  sufficiently large. Stochastic orbits enable ions to absorb energy
  from the turbulence, increasing the perpendicular ion
  temperature~$T_{\perp \rm i}$ even when the fluctuation frequencies
  are too small for a cyclotron resonance to occur.  In this paper, an
  analytic expression for the stochastic heating rate is used in
  conjunction with an observationally constrained turbulence model to
  obtain an analytic formula for $T_{\perp \rm i}$ as a function of
  heliocentric distance~$r$, ion mass, and ion charge in coronal holes
  at $2 R_{\sun} \lesssim r < 15 R_{\sun}$.  The resulting temperature
  profiles provide a good fit to observations of protons and~${\rm
    O}^{+5}$ ions at $2 R_{\sun} \lesssim r \lesssim 3 R_{\sun}$ from
  the {\em Ultraviolet Coronagraph Spectrometer} (UVCS).  Stochastic
  heating also offers a natural explanation for several detailed
  features of the UVCS observations, including the preferential and
  anisotropic heating of minor ions, the rapid radial increase in
  the~${\rm O}^{+5}$ temperature between~$1.6 R_{\sun}$ and~$1.9
  R_{\sun}$, and the abrupt flattening of the ${\rm O}^{+5}$
  temperature profile as $r$ increases above~$1.9 R_{\sun}$.
\end{abstract}

\maketitle

\vspace{0.2cm} 
\section{Introduction}
\label{sec:intro}
\vspace{0.2cm} 

Measurements of ion and electron temperatures show that the solar wind
undergoes substantial heating throughout its transit from the solar
surface to the heliospheric termination shock. A number of
considerations suggest that much of this heating results from
Alfv\'en-wave turbulence, as suggested over forty years ago
by~\cite{coleman68}. For example, Alfv\'en-wave-like fluctuations in
the velocity, magnetic field, and electric field are often present in
the interplanetary medium, with power spectra that vary approximately
as power laws over a broad range of length
scales~\citep{belcher71,goldstein95a,tumarsch95,bruno05,bale05}.
Alfv\'en waves have also been measured remotely in the
chromosphere~\citep{depontieu07} and corona~\citep{tomczyk07}, and
radio observations at heliocentric distances~$r$ as small as~$\sim
5R_{\sun}$ reveal a broad spectrum of density fluctuations that are
consistent with passive-scalar mixing of entropy modes by
Alfv\'en-wave turbulence~\citep{harmon05,chandran09d}.  At $r\gtrsim
60 R_{\sun}$, the amplitudes of velocity and magnetic-field
fluctuations are positively correlated with
temperature~\citep{grappin90}, and the turbulent dissipation rate
implied by the measured fluctuation amplitudes is comparable to the
heating rate inferred from measurements of the proton and electron
temperature profiles~\citep{smith01,cranmer09,stawarz09}.

On the other hand, it has for some time been unclear whether
Alfv\'en-wave turbulence can explain the ion temperatures and
temperature anisotropies observed in the fast solar wind and coronal
holes (the open-magnetic-field-line regions from which the fast wind
emanates).  For example, observations from the {\em Ultraviolet
  Coronagraph Spectrometer}~(UVCS) show that minor ions such as~${\rm
  O}^{+5}$ are heated to temperatures greatly exceeding the proton and
electron temperatures at $2 R_{\sun} \lesssim r \lesssim 3 R_{\sun}$,
with thermal motions that are much more rapid in directions
perpendicular to the magnetic field~${\bf B}$ than along the magnetic
field --- i.e., $T_\perp \gg T_\parallel$~(Kohl et al.~1998; Li et
al.~1998; Antonucci et al.~2000). \nocite{kohl98,li98,antonucci00}
Similarly, in-situ measurements from the Helios and Wind satellites
show that $T_{\perp \rm p} > T_{\parallel \rm p} > T_{\rm e}$ in
low-$\beta$ fast-wind streams at $60 R_{\sun} \lesssim r \leq
\mbox{1~AU}$, where $T_{\perp \rm p}$ and $T_{\parallel \rm p}$ are
the perpendicular and parallel proton temperatures, $T_{\rm e}$ is the
electron temperature, and $\beta = 8\pi p/B^2$ is the ratio of the
plasma pressure~$p$ to the magnetic
pressure~\citep{phillips90,marsch04}.

These observations lead to an apparent difficulty for
solar-wind-heating models based on Alfv\'en-wave turbulence for the
following reasons.  In Alfv\'en-wave turbulence, the energy cascade is
anisotropic, transporting energy primarily to small scales measured
perpendicular to~${\bf B}$ rather than small scales along~${\bf
  B}$~\citep{shebalin83,goldreich95,ng96,galtier00,cho02b}.  In
wavenumber space, energy cascades primarily to larger~$k_\perp$, and
only weakly to larger~$k_\parallel$, where $k_\perp$ and $k_\parallel$
are the components of the wave vector~${\bf k}$ perpendicular and
parallel to~${\bf B}$.  At perpendicular scales~$\lambda_\perp$ of
order the proton gyroradius~$\rho_{\rm p}$, the Alfv\'en-wave cascade
transitions to a kinetic-Alfv\'en-wave (KAW)
cascade~\citep{bale05,howes08b,schekochihin09,sahraoui09}, and at
$\lambda_\perp \lesssim \rho_{\rm p}$ the fluctuations dissipate.  The
rms amplitude of the magnetic-field fluctuations at~$\lambda_\perp
\lesssim \rho_{\rm p}$ is~$\ll B$, and thus it seems reasonable to
assume that the KAW fluctuations damp at the same rate as linear KAWs
at the same~${\bf k}$.  For the~$\beta$ values found in coronal holes
and the solar wind, linear KAWs undergo significant electron Landau
damping~\citep{quataert98,leamon99}.  However, the KAWs produced by
the anisotropic cascade do not undergo ion cyclotron damping, because
their frequencies~$\omega$ are much less than the proton cyclotron
frequency~\citep{cranmer03,howes08a}.  In addition, when~$\beta \ll
1$, the ion thermal speeds are~$\ll v_{\rm A}$, where $v_{\rm A}$ is
the Alfv\'en speed.  Ions are thus unable to satisfy the resonance
condition~$\omega - k_\parallel v_\parallel = 0 $ for Landau damping
or transit-time damping, where $v_\parallel$ is the velocity component
parallel to~${\bf B}$, because $\omega/k_\parallel \geq v_{\rm A}$ for
KAWs~\citep{quataert98,hollweg99c}. As a consequence, linear damping
of KAWs by ions is negligible when~$\beta \ll 1$, suggesting that
low-frequency Alfv\'en-wave/KAW turbulence results in negligible ion
heating in coronal holes and low-$\beta$ fast-wind streams.

A number of studies have gone beyond the framework of linear wave
theory to explore the possibility of perpendicular ion heating by
low-frequency turbulence in the solar wind (e.g., Dmitruk et al.~2004;
Markovskii et al.~2006; Bourouaine et al.~2008; Parashar et al.~2009;
Lehe et al.~2009; -- see also Drake et al.~2009).
\nocite{dmitruk04,markovskii06,bourouaine08,parashar09,lehe09,drake09} Several
investigations have shown that low-frequency Alfv\'en waves and KAWs
can indeed cause perpendicular ion heating if the wave amplitudes are
sufficiently large~\citep{johnson01,chen01,white02,voitenko04}. This
type of heating is some times referred to as ``stochastic heating,''
since it is predicated upon the stochastic ion orbits that arise in
the presence of large-amplitude fluctuations in the electric or
magnetic field. This paper addresses the question of whether
stochastic ion heating can explain the anisotropic ion temperatures
that are observed in coronal holes. The analysis is based on the
recent studies by \cite{chandran10} and \cite{chandran09c}, which are
reviewed in sections~\ref{sec:heating} and~\ref{sec:ref}. Stochastic
heating in coronal holes and the inner solar wind is then discussed
section~\ref{sec:corona}.

\vspace{0.2cm} 
\section{Stochastic Ion Heating}
\label{sec:heating} 
\vspace{0.2cm} 

If the amplitudes of Alfv\'en waves or KAWs with $\lambda_\perp \sim
\rho_{\rm i}$ and $\omega < \Omega_{\rm i}$ are sufficiently large,
then ions undergo ``stochastic heating,'' where $\lambda_\perp$ is the
wavelength measured perpendicular to~${\bf B}$, $\omega$ is the wave
frequency, $\rho_{\rm i} = v_{\perp\rm i}/\Omega_{\rm i}$ is the ion
gyroradius, $v_{\perp \rm i} = \sqrt{2k_{\rm B} T_{\perp \rm i}/
  m_{\rm i}}$ is the rms value of~$v_\perp$ (the ion velocity
component perpendicular to~${\bf B}$ in the local plasma frame),
$\Omega_{\rm i} = q_{\rm i} B/m_{\rm i} c$ is the ion cyclotron
frequency, and $m_{\rm i}$, $q_{\rm i}$ and $T_{\perp \rm i}$ are the
ion mass, charge, and perpendicular
temperature~\citep{johnson01,chen01}. \cite{chandran10} showed that
the amplitude threshold for strong stochastic heating when~$\beta
\lesssim 1$ can be expressed in terms of the quantity
\begin{equation}
\epsilon_{\rm i}= \frac{\delta v_{\rm i}}{v_{\perp \rm i}},
\label{eq:defeps} 
\end{equation} 
where $\delta v_{\rm i}$ is the rms amplitude of the velocity
fluctuation at $\lambda_\perp \sim \rho_{\rm i}$.  For Alfv\'en-wave
and KAW fluctuations, $\epsilon_{\rm i} \sim q_{\rm i} \delta
\Phi_{\rm i}/m v_{\perp \rm i}^2$, where $\delta \Phi_{\rm i}$ is the
rms amplitude of the electrostatic-potential fluctuation at
$\lambda_\perp \sim \rho_{\rm i}$, and thus $\epsilon_{\rm i}$ is roughly
the fractional change in an ion's kinetic energy induced by
gyroscale fluctuations during the course of a single cyclotron period.
If $\epsilon_{\rm i}$ is sufficiently small, then a thermal ion's
orbit in the plane perpendicular to~${\bf B}$ closely approximates a
closed circle in some suitably chosen reference frame. In this case,
the ion's magnetic moment $\mu = m_{\rm i} v_\perp^2/2B$ is nearly
conserved, and perpendicular ion heating is weak~\citep{kruskal62}. On
the other hand, as $\epsilon_{\rm i}$ increases towards unity, a
thermal ion's orbit becomes increasingly chaotic, $\mu$ conservation
breaks down, and perpendicular ion heating becomes increasingly
strong. For protons, $v_{\perp \rm i} \sim \beta^{1/2} v_{\rm A}$,
$\delta v_{\rm i} \sim v_{\rm A} \delta B_{\rm i}/B$, and
$\epsilon_{\rm i} \sim \beta^{-1/2} \delta B_{\rm i}/B$, where $\delta
B_{\rm i}$ is the rms amplitude of the magnetic fluctuation at
$\lambda_{\perp} \sim \rho_{\rm i}$. Thus, when $\beta \ll 1$,
$\epsilon_{\rm i}$ is much greater than~$\delta B_{\rm i}/B$.

Using phenomenological arguments, \cite{chandran10} derived an
expression (their equation (37)) for the time scale~$t_{\rm h}$ on
which stochastic heating doubles an ion's kinetic energy in the
absence of cooling:
\begin{equation}
  t_{\rm h} \simeq \epsilon_{\rm i}^{-3}\Omega_{\rm i}^{-1} \exp \left( \frac{c_2}{\epsilon_{\rm i}}\right),
\label{eq:th1} 
\end{equation} 
where $c_2$ is a dimensionless constant. An important consequence of
equation~(\ref{eq:th1}) is that stochastic heating is inherently
self-limiting --- as $T_{\perp \rm i}$ grows, $\epsilon_{\rm i}$
decreases, and the stochastic heating rate decreases as a result.
\cite{chandran10} evaluated $c_2$ by simulating test-particle protons
interacting with a spectrum of randomly phased Alfv\'en waves and KAWs
in a low-$\beta$ plasma and found that~$c_2 = 0.34$. However, they
argued that $c_2$ is smaller for Alfv\'en-wave/KAW turbulence than for
randomly phased waves, because turbulence produces coherent structures
that increase orbit stochasticity~\citep{dmitruk04}.  In the present
paper, $c_2$ is taken to be the same for all ion species, and the
value
\begin{equation}
c_2 = 0.15 
\label{eq:valc2} 
\end{equation} 
is chosen in order to match UVCS observations of ${\rm O}^{+5}$
temperatures, as discussed further in section~\ref{sec:corona} below.
\cite{chandran10} also showed that in low-$\beta$ plasmas stochastic
heating by low-frequency Alfv\'en-wave/KAW turbulence primarily
increases an ion species' perpendicular temperature~$T_{\perp \rm i}$,
rather than its parallel temperature~$T_{\parallel \rm i}$.

\vspace{0.2cm} 
\section{Alfv\'en-Wave Turbulence in Coronal Holes}
\label{sec:ref} 
\vspace{0.2cm} 

Convective motions at the surface of the Sun randomly stir the
footpoints of open magnetic field lines, launching low-frequency
Alfv\'en waves that propagate into coronal holes and then on into the
solar wind. Radial variations in the Alfv\'en speed~$v_{\rm A}(r)$
couple the outward-propagating Alfv\'en waves to inward-propagating
Alfv\'en waves~\citep{heinemann80,velli93,hollweg07,cranmer10}, and
nonlinear interactions among counter-propagating waves then cause the
wave energy to cascade to small scales and dissipate, heating the
ambient plasma~\citep{velli89,matthaeus99b,cranmer07,verdini10}.  The
value of the correlation length or ``outer scale'' at the coronal
base, denoted $L_{\perp\,\rm b}$, is in the present study taken to be
$5\times 10^3$~km, comparable to the spacing of photospheric flux
tubes~\citep{spruit81}. At larger~$r$, the value of~$L_\perp$ is taken
to increase in proportion to the radius of a flux tube, so that
\begin{equation}
L_\perp(r) = L_{\perp\,\rm b}\left[\frac{B_0(r_{\rm b})}{B_0(r)}\right]^{1/2},
\label{eq:defLperp} 
\end{equation} 
where $B_0$ is the background magnetic field strength and $r_{\rm b}
\simeq 1 R_{\sun}$ corresponds to the coronal base (i.e., $x=1$ in
equation~(\ref{eq:B0}) below).

\cite{chandran09c} developed an analytical model of Alfv\'en-wave
reflection and turbulent heating based upon a phenomenological
treatment of the energy cascade. Using this model, they calculated the
radial profiles of the turbulent dissipation rate per unit
mass~$\Gamma$ and rms wave amplitude at the outer scale,
denoted~$\delta v_0$, extending the previous model of \cite{dmitruk02}
to account for the solar-wind velocity. For the case that most of the
Alfv\'en-wave energy is at periods of $\sim 1$~hour or longer,
\cite{chandran09c} found that 
\begin{equation}
\delta v_0  = \left( \frac{2\eta^{1/4}}{ 1 + \eta^{1/2}}\right)
\left(\frac{v_{\rm A}}{v_{\rm Aa}} \right)^{\!\!\! 1/2}\delta v_{\rm a}
\label{eq:dv0b} 
\end{equation} 
and
\begin{equation}
\Gamma = (1+\eta^{-1/2})\,\delta v_0^2 \,  \left|\frac{dv_{\rm A}}{dr}\right|
\label{eq:Gamma0} 
\end{equation} 
for $r_{\rm m} < r \lesssim 20 R_{\sun}$, where $r_{\rm m}$ is the
radius at which the Alfv\'en speed $v_{\rm A} = B_0/\sqrt{4\pi n
  m_{\rm p}}$ obtains its maximum value, $n$ is the proton number
density, $m_{\rm p}$ is the proton mass, $\delta v_{\rm a}$ is the value of $\delta v_0$ at $r =
r_{\rm a}$, $r_{\rm a}$ is the radius of the Alfv\'en critical point
at which the solar-wind outflow speed~$U$ equals~$v_{\rm A}$, $v_{\rm
  Aa}$ is the Alfv\'en speed at $r=r_{\rm a}$, and
\begin{equation}
\eta(r) = \frac{n(r)}{n(r_{\rm a})}.
\label{eq:defeta} 
\end{equation} 

\begin{figure}[t]
\centerline{\includegraphics[width=8.cm]{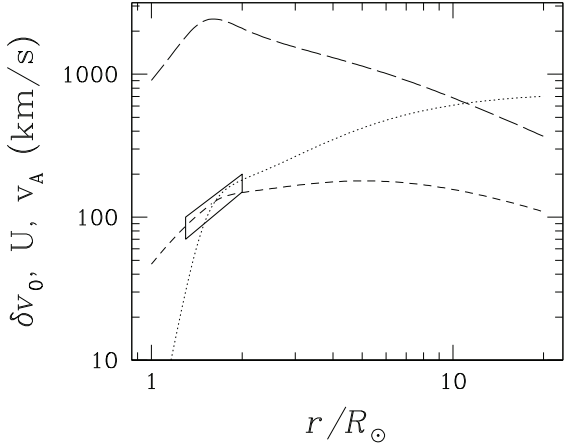}}
\caption{The short-dashed line is the rms amplitude of the fluctuating
  velocity from equation~(\ref{eq:dv0b}) with $\delta v_{\rm a} =
  155$~km/s, where $n$, $B_0$, and $U$ are taken from
  equations~(\ref{eq:n}) through (\ref{eq:defU}). The solid-line box
  represents the range of non-thermal line widths
  obtained by \cite{esser99}. The dotted line is the solar-wind velocity from
  equation~(\ref{eq:defU}), and the long-dashed line is the Alfv\'en speed from
  equations~(\ref{eq:n}) and (\ref{eq:B0}).
  \label{fig:swv_corona}}\vspace{0.5cm}
\end{figure}

The above results can be applied to coronal holes and the fast solar
wind with the use of the following model profiles for $n$, $B_0$, and~$U$:
\begin{equation}
n = \left(\frac{3.23 \times 10^8}{x^{15.6}} + \frac{2.51 \times 10^6}{x^{3.76}} + \frac{1.85 \times 10^5}{x^2}\right) \mbox{ cm}^{-3},
\label{eq:n} 
\end{equation} 
\begin{equation}
B_0 = \left[\frac{6}{x^6} + \frac{1.5}{x^2}\right] \mbox{ Gauss},
\label{eq:B0} 
\end{equation} 
and
\begin{equation}
U = 9.25 \times 10^{12} \;\frac{\tilde{B}}{\tilde{n}} \;\mbox{cm}\;\mbox{s}^{-1} ,
\label{eq:defU} 
\end{equation} 
where $x = r/R_{\sun}$, $\tilde{B}$ is $B_0$ in Gauss, and $\tilde{n}$
is $n$ in units of $\mbox{cm}^{-3}$.  The density in
equation~(\ref{eq:n}) is the value from equation (4) of
\cite{feldman97} plus an additional~$r^{-2}$ component chosen to give
$n=4 \mbox{ cm}^{-3}$ at 1~AU.  Equations~(\ref{eq:n}) through
(\ref{eq:defU}) give $r_a=11.1 R_{\sun}$, $r_m= 1.60 R_{\sun}$,
 and $U(\mbox{1 AU}) = 750\;
\mbox{km}\; \mbox{s}^{-1}$, and lead to the $U$ and $v_{\rm A}$
profiles shown in figure~\ref{fig:swv_corona}.  The value of $\delta
v_0$ from equation~(\ref{eq:dv0b}) is also plotted in
figure~\ref{fig:swv_corona}, where the value $\delta v_{\rm a} =
155$~km/s has been chosen so that $\delta v_0$ remains bounded by the
range of  non-thermal line widths
obtained by \cite{esser99} multiplied
by~$\sqrt{2}$ to convert from an rms line-of-sight velocity to an rms
velocity in the plane perpendicular to~${\rm B}$.

In Alfv\'en-wave turbulence, fluctuation energy cascades from
$\lambda_\perp \sim L_\perp$ to smaller~$\lambda_\perp$ and then
dissipates at $\lambda_\perp \lesssim \rho_{\rm p}$ due to some
combination of ion and electron damping. It is assumed that
dissipation is negligible at $\lambda_\perp > 2\rho_{\rm p }$, and
that
\begin{equation}
  \delta v_{\lambda_\perp } = \delta v_0 \left(\frac{\lambda_\perp}{L_\perp}\right)^a
\label{eq:spectrum} 
\end{equation} 
at $2\rho_{\rm p } < \lambda_\perp < L_\perp$, where $\delta
v_{\lambda_\perp}$ is the rms amplitude of the fluctuating velocity at
perpendicular scale~$\lambda_\perp$. The constant~$a$ is related to
the velocity power spectrum~$P_{\rm v}(k_\perp)$. If $P_{\rm v}\propto
k_\perp^{-\sigma}$ for $ L_\perp^{-1} \lesssim k_\perp \lesssim
(2\rho_{\rm p})^{-1}$, then $a = (\sigma-1)/2$. Observations of
solar-wind velocity fluctuations at 1~AU find that~$\sigma =
3/2$~\citep{podesta07}.  Numerical simulations of magnetohydrodynamic
(MHD) turbulence generally find that~$\sigma = 5/3$ when~$\delta v_0
\sim v_{\rm A}$~\citep{cho00,muller00,haugen04} and $\sigma = 3/2$
when~$\delta v_0\lesssim 0.2 v_{\rm
  A}$~\citep{maron01,muller05,boldyrev06,mason08,perez08a,perez09a}.
From figure~\ref{fig:swv_corona}, $\delta v_0/v_{\rm A} < 0.3$ at $r<
15 R_{\sun}$, indicating that this near-Sun region is better described
by simulations with $\delta v_0 \lesssim 0.2 v_{\rm A}$ than
simulations with~$\delta v_0 \sim v_{\rm A}$. Based on the above
studies, it is assumed that~$\sigma = 3/2$ and
\begin{equation}
a= 0.25
\label{eq:vala} 
\end{equation} 
at $r \leq 15 R_{\sun}$, the region on which this study focuses.

UVCS observations show that ${\rm O}^{+5}$ has a larger thermal speed
than protons at $r \sim 2 R_{\sun}$, and hence a gyroradius that is
several times larger than~$\rho_{\rm p}$. Based in part on this
observation, it is assumed that $\rho_{\rm i} > 2 \rho_{\rm p}$ for
minor ions and alpha particles at~$r \gtrsim 2 R_{\sun}$.  Equation~(\ref{eq:spectrum})
then gives
\begin{equation}
  \delta v_{\rm i } = \alpha_{\rm i}\delta v_0 \left(\frac{\rho_{\rm i}}{L_\perp}\right)^a,
\label{eq:spectrumi} 
\end{equation} 
with 
\begin{equation}
\alpha_{\rm i} = 1 \hspace{0.5cm} \mbox{(for ${\rm He}^{++}$ and minor ions)}.
\label{eq:alpha1} 
\end{equation} 
In the case of protons, electron Landau damping and stochastic proton
heating drain energy from the cascade at $\lambda_\perp \sim \rho_{\rm
  p}$, reducing $\delta v_{\lambda_\perp}$ at $\lambda_\perp \sim
\rho_{\rm p}$ relative to the scaling in
equation~(\ref{eq:spectrum}). To account for this, it is
assumed that
\begin{equation}
\alpha_{\rm i} = 0.71 \hspace{0.5cm} \mbox{(for protons)},
\label{eq:alpha2} 
\end{equation} 
where the particular value in equation~(\ref{eq:alpha2}) is chosen so
that the results in section~\ref{sec:corona} match observations
of~$T_{\perp \rm p}$ in coronal holes, as described further  in
the discussion of figure~\ref{fig:Tcomp}.
Equations~(\ref{eq:defeps}) and (\ref{eq:spectrumi}) imply that
\begin{equation}
\epsilon_{\rm i} = \alpha_{\rm i} \left(\frac{\delta v_0}{v_{\perp \rm i}}\right)^{1-a}
\left(\frac{\delta v_0}{L_\perp \Omega_{\rm i}}\right)^a.
\label{eq:eps2} 
\end{equation} 
Equation~(\ref{eq:eps2}) shows that as~$T_{\perp \rm i}$ decreases,
$\epsilon_{\rm i}$ increases, which in turn increases the stochastic
heating rate.  One way of understanding this is that Alfv\'en-wave/KAW
fluctuations with $\lambda_\perp \sim \rho_{\rm i}$ cause the
electrostatic potential~$\Phi$ to vary in a complicated way in the
plane perpendicular to~${\bf B}_0$, with an rms variation of $\delta \Phi_{\rm i}
\sim \rho_{\rm i} \delta E_{\rm i}$ over a distance~$\rho_{\rm i}$,
where $\delta E_{\rm i} \sim \delta v_{\rm i } B_0/c$ is the rms
amplitude of the electric-field fluctuation at $\lambda_\perp =
\rho_{\rm i}$. (The larger but smoother spatial variations in~$\Phi$
associated with Alfv\'en-wave fluctuations at~$\lambda_\perp \gg
\rho_{\rm i}$ are ignored here, as they lead to drift motion rather
than stochastic orbits.)  Equation~(\ref{eq:spectrumi}) then gives
$\delta \Phi_{\rm i} \propto v_{\perp \rm i}^{1+a}$, and $\delta
\Phi_{\rm i}/ v_{\perp \rm i}^2 \propto v_{\perp \rm i}^{ a-1}$. For
$a< 1$, decreasing $v_{\perp \rm i}$ causes $q_{\rm i}\delta \Phi_{\rm
  i}$ to become an increasingly large fraction of the ions'
perpendicular kinetic energy, and the ions' motion in the plane
perpendicular to~${\bf B}_0$ becomes increasingly chaotic as a result.

A possible objection to setting $\rho_{\rm i} > 2 \rho_{\rm p}$ at
$r\gtrsim 2 R_{\sun}$ for ${\rm He}^{++}$ and minor ions is that
preferential heating of heavy ions is assumed from the outset, leaving
open the question of how such ions first reach temperatures
exceeding~$T_{\perp \rm p}$.  This initial evolution can be understood
by considering ${\rm He}^{++}$, ${\rm O}^{+5}$, and ${\rm Fe}^{+11}$
ions with temperatures~$\sim T_{\perp \rm p}$ at small~$r$.  Given
their charge-to-mass ratios, these ions have gyroradii that are
comparable to~${\rm \rho}_{\rm p}$ when $T_{\perp \rm i} \sim T_{\perp
  \rm p}$, and hence values of $\delta v_{\rm i}$ that are comparable
to the proton value.  As a result, the slower, heavier ions have
larger~$\epsilon_{\rm i}$ and much larger stochastic heating rates
than protons when~$T_{\perp \rm i} \sim T_{\perp \rm p}$. These larger
heating rates then lead to $T_{\perp \rm i} \gg T_{\perp \rm p}$ at
larger~$r$.

In the {\em Cluster} measurements at 1~AU analyzed by \cite{bale05}, the
electric-field power spectrum is slightly larger at $k_\perp \rho_{\rm
  p} = 1$ than one would expect from an extrapolation of the power-law
scaling that is present at smaller~$k_\perp$.  The electric-field
spectrum is a good proxy for the spectrum of (electron) velocity
fluctuations associated with AW/KAW turbulence~\citep{schekochihin09},
and thus $\alpha_{\rm i} > 1$ for protons in this data set.  Nevertheless,
the value~$\alpha_{\rm i} = 0.71$ is reasonable for protons in coronal
holes for two reasons. First, AW/KAW turbulence is more ``imbalanced''
in coronal holes than at 1~AU, with a greater excess of
outward-propagating waves over inward propagating
waves~\citep{cranmer05,verdini07}. Such imbalance weakens the cascade
rate, causing the power spectrum to decrease more rapidly with
increasing~$k_\perp$ near $k_\perp \rho_{\rm p} = 1$ in response to a
fixed KAW damping rate. Second, the linear damping rate of KAWs at
$k_\perp \rho_{\rm p} = 1$ is significantly larger in the $\beta \ll
1$ conditions of coronal holes than in the typical $\beta \sim 0.5 -
1$ conditions found at 1~AU (see, e.g., figure~2 of \cite{howes08a}).

\vspace{0.2cm} 
\section{Ion Temperature Profiles}
\label{sec:corona} 
\vspace{0.2cm} 

As described in the introduction, linear damping of KAWs by ions is
extremely weak when~$\beta \ll 1$~\citep{quataert98}.  Based in part
on this finding, it is assumed in this section that stochastic heating
is the primary ion heating mechanism in coronal holes. The $T_{\perp
  \rm i}$ profiles that result from this assumption are then
calculated, and it is shown that for plausible parameter values
the resulting profiles provide a good fit to the observations.

At $r\gtrsim 2 R_{\sun}$, collisional energy exchange
between particle species can be neglected to a reasonable
approximation, because the energy exchange
time scale exceeds the expansion time scale~\citep{esser99},
\begin{equation}
t_{\rm exp } = \frac{r}{U_{\rm i}}.
\label{eq:texp} 
\end{equation} 
The time scale~$t_{\rm cond}$ on which ion temperatures are modified
by ion thermal conduction is $\gtrsim r/v_{\parallel \rm i}$, where
this lower limit corresponds to ions streaming freely at their parallel
thermal velocity $v_{\parallel \rm i} = \sqrt{k_{\rm B} T_{\parallel
    \rm i}/m_{\rm i}}$. At $r\gtrsim 2 R_{\sun}$, ion flows are
supersonic~\citep{kohl98}, and thus $t_{\rm cond}> t_{\rm exp}$ even
in the free-streaming limit. Ion thermal conduction can thus be
neglected at $r\gtrsim 2 R_{\sun}$ to a first approximation.

Because collisions and conduction are weak, minor ions at
$r\gtrsim 2 R_{\sun}$ evolve towards a state in which
\begin{equation}
t_{\rm h} \sim t_{\rm exp}
\label{eq:thtexp} 
\end{equation} 
for the following reasons. Minor ions
draw very little power from the turbulence, and the power spectrum of
the turbulence is essentially independent of the minor-ion heating
rate~$Q_{\rm i}$. Equations~(\ref{eq:defeps}) and (\ref{eq:th1}) thus
imply that $Q_{\rm i}$ is a highly sensitive function of the minor-ion
temperature, with strong heating at small~$T_{\perp \rm i}$ and
exponentially weak heating at sufficiently large~$T_{\perp \rm i}$.
If $t_{\rm h}\gg t_{\rm exp}$ at some radius~$r_1$, then adiabatic
cooling would cause~$T_{\perp \rm i}$ to decrease in proportion
to~$B_0$, and $d\ln T_{\perp \rm i}/d\ln r$ would be~$\leq -2$
near~$r_1$.  The radial decrease in $T_{\perp \rm i}$ near~$r_1$ would
lead to a sharp decrease in $t_{\rm h}$, and at larger~$r$ the minor
ions would leave the adiabatic regime.  Conversely, if $t_{\rm h}\ll
t_{\rm exp}$ at some radius~$r_2$, then $d\ln T_{\perp \rm i}/d\ln r
\gg 1$ near~$r_2$. The radial increase in~$T_{\perp \rm i}$ would
cause $t_{\rm h}$ to increase rapidly with increasing~$r$, and at
larger~$r$ the ions would again approach a state in which~$t_{\rm h}
\sim t_{\rm exp}$.

The observed $T_{\perp \rm i}$ profile of~${\rm O}^{+5}$ ions in
coronal holes (see figure~\ref{fig:Tcomp} below) appears to provide an
example of the second case just described, in which ions swiftly
evolve from a state in which $t_{\rm h} \ll t_{\rm exp}$ to a state in
which~$t_{\rm h} \sim t_{\rm exp}$.  The ${\rm O}^{+5}$ temperature is
$ <10^7$~K at~$r\leq 1.6 R_{\sun}$, presumably due to collisions with
protons.  As ${\rm O}^{+5}$ ions flow outward past $r=1.6 R_{\sun}$,
their temperature increases rapidly to $\sim 10^8$~K at $r= 1.9
R_{\sun}$, where collisional energy exchange with protons is weak.  At
$r = 1.9 R_{\sun}$, the ${\rm O}^{+5}$ temperature profile abruptly
flattens, with $T_{\perp \rm i}$ remaining fairly constant out to~$2.7
R_{\sun}$, indicating that~$t_{\rm h} \sim t_{\rm exp}$ at $r\gtrsim 2
R_{\sun}$.  These rapid radial variations in $T_{\perp \rm i}$ and
$dT_{\perp \rm i}/dr$ present a challenge for theoretical models, but
can be naturally explained in terms of stochastic heating ---
stochastic heating of an initially cool minor-ion population quickly
increases $T_{\perp \rm i}$, but then saturates at large~$T_{\perp \rm
  i}$ because of the reduction in orbit stochasticity.\footnote{It
  should be noted that the observed ${\rm O}^{+5}$ temperature profile
  has also been approximately reproduced by models invoking resonant
  cyclotron heating by high-frequency Alfv\'en/ion-cyclotron waves ---
  see, e.g., \cite{cranmer99} and~\cite{isenberg09}.}

\begin{figure}[t]
\centerline{\includegraphics[width=8.cm]{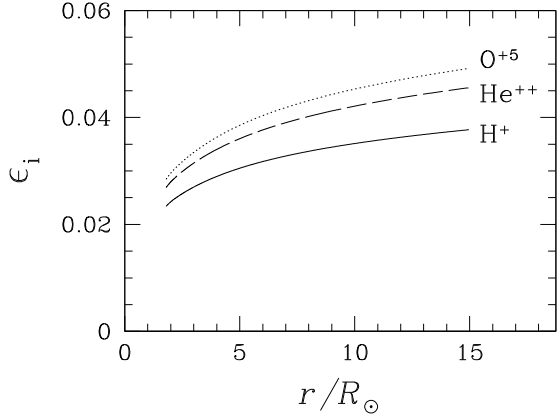}}
\caption{The values of $\epsilon_{\rm i}$ from equation~(\ref{eq:th2})
  for protons, alpha particles, and ${\rm
    O}^{+5}$ ions, where $c_2 = 0.15$ and $\Omega_{\rm
    i}(r)$ is calculated using equation~(\ref{eq:B0}). For protons,
  $U_{\rm i}$ in equation~(\ref{eq:th2}) is set equal to the value of
  $U$ in equation~(\ref{eq:defU}). For ${\rm He}^{++}$ and ${\rm
    O}^{+5}$, $U_{\rm i}$ is set equal to $1.75 U$.
  \label{fig:eps_profile}}\vspace{0.5cm}
\end{figure}

The strong dependence of~$t_{\rm h}$ on~$\epsilon_{\rm i}$ implies that
$t_{\rm h} \sim t_{\rm exp}$ only within a narrow interval of $\epsilon_{\rm
  i}$ values. The midpoint of this interval can be found by equating
$t_{\rm h} $ and $t_{\rm exp}$, which yields
\begin{equation}
 \epsilon_{\rm i}^{-3} \exp \left( \frac{c_2}{\epsilon_{\rm i}}\right)
=  \frac{ \Omega_{\rm i} r}{U_{\rm i}}.
\label{eq:th2}
\end{equation} 
The right-hand side of this equation is~$\gg 1$ in coronal holes,
leading to values of~$\epsilon_{\rm i}$ that are $ \ll 1$.  For
example, $\Omega_{\rm i} r/U_{\rm i} = 6.13 \times 10^6$ for ${\rm
  O}^{+5}$ ions at~$r=2 R_{\sun}$ given the assumptions listed in the
caption of figure~\ref{fig:eps_profile}. Equation~(\ref{eq:th2})
then gives $\epsilon_{\rm i} = 2.96\times 10^{-2}$, assuming~$c_2 =
0.15$. For such small values of~$\epsilon_{\rm i}$, the value
of~$\epsilon_{\rm i}$ becomes relatively insensitive to changes in the
right-hand side of equation~(\ref{eq:th2}). For example, if
$\Omega_{\rm i} r/U_{\rm i}$ is increased from $6.13 \times 10^6$ by
50\%, the resulting decrease in~$\epsilon_{\rm i}$ is only~5\%.
 
For protons, the comparatively flat $T_{\perp \rm i}$ profiles seen in
UVCS observations (Kohl et al.~1998) rule out the possibility that
$t_{\rm h} \ll t_{\rm exp}$ or $t_{\rm h} \gg t_{\rm exp}$, assuming
stochastic heating is the dominant heating mechanism. \nocite{kohl98}
Thus, $t_{\rm h} \sim t_{\rm exp}$ and protons approximately satisfy
equation~(\ref{eq:th2}).  However, protons can attain the required
value of~$\epsilon_{\rm i}$ not only by getting hotter or cooler, but
also by absorbing energy from the turbulence and reducing the value
of~$\delta v_{\lambda_\perp}$ at~$\lambda_\perp \sim \rho_{\rm p}$.
No attempt is made in this paper to treat the coupled evolution of
protons and gyro-scale KAW fluctuations self-consistently. Instead,
proton damping (and electron Landau damping) of fluctuations
at~$\lambda_\perp \sim \rho_{\rm p}$ is modeled simplistically by
setting~$\alpha_{\rm i} = 0.71$ for protons in
equation~(\ref{eq:alpha2}). This particular value is chosen to match
the UVCS observations shown in figure~\ref{fig:Tcomp}, as described
further below.  Although helium comprises only $\sim 5\%$ of the ions
in the fast solar wind~\citep{bame77}, alpha particles are hotter than
protons in the fast wind, and may also drain a significant amount of
power from the cascade~\citep{marsch82a,kasper07}.\footnote{For
  example, the Helium heating rate is comparable to the proton heating
  rate for the temperature profiles in the right panel of
  figure~\ref{fig:Tcomp} given the assumed values of~$U_{\rm i}$.} The
back reaction of helium heating upon the turbulence, however, is
neglected in this paper.

To determine~$\epsilon_{\rm i}$ from equation~(\ref{eq:th2}), $B_0$
is taken from equation~(\ref{eq:B0}) and $U_{\rm i}$ for protons is
set equal to~$U$ in equation~(\ref{eq:defU}). For other ion species,
$U_{\rm i}$ is taken to be~1.75 times the proton speed, which is
consistent with UVCS observations of protons and~${\rm O}^{+5}$ ions
at~$r=3 R_{\sun}$ (Kohl et al.~1998), \nocite{kohl98} but is only a
rough approximation for other ion species and at other radii.  The
resulting values of~$\epsilon_{\rm i}$ for protons, alpha particles,
and~${\rm O}^{+5}$ ions are shown in figure~\ref{fig:eps_profile} for
$ 1.8 R_{\sun}< r < 15 R_{\sun}$. Since ion thermal conduction and
collisional energy exchange between particle species are neglected,
the curves are not extended to smaller~$r$ where these processes
become important.

Once $\epsilon_{\rm i}(r)$ is determined using
equation~(\ref{eq:th2}), $T_{\perp \rm i}$ can be determined from
equation~(\ref{eq:eps2}), which can be re-written as
\begin{equation}
T_{\perp \rm i } = \frac{ m_{\rm i}}{2 k_{\rm B}} \left[\frac{\alpha_{\rm i} \delta v_0}{\epsilon_{\rm i} (L_\perp \Omega_{\rm i})^a}\right]^{2/(1-a)}.
\label{eq:Tsat} 
\end{equation} 
As illustrated in figure~\ref{fig:eps_profile},
equation~(\ref{eq:th2}) leads to similar values of~$\epsilon_{\rm i}$
for alpha particles and minor ions. If $\epsilon_{\rm i}$ and
$\alpha_{\rm i}$ are the same for some set of ion species (or if
$\alpha_{\rm i}/\epsilon_{\rm i}$ is the same), then
equation~(\ref{eq:Tsat}) implies that
\begin{equation}
T_{\perp \rm i} \propto   A \left(\frac{A}{Z}\right)^{l} 
\label{eq:Tscaling} 
\end{equation} 
for these ions,
where $A = m_{\rm i}/m_{\rm p}$, $Z = q_{\rm i}/e$, $e$ is the proton charge,
and $l = 2a/(1-a)$. The value of~$l$ can also be expressed as
\begin{equation}
l = \frac{2\sigma - 2}{3 - \sigma},
\label{eq:defl} 
\end{equation} 
where $\sigma $ is the spectral index of the velocity power spectrum
defined in section~\ref{sec:heating}. In this paper, it is assumed
that $a = 1/4$ and $\sigma = 3/2$, which gives $l= 2/3$.  If instead $
a= 1/3$ and $\sigma = 5/3$, then $l = 1$.  

The largest radius at which equations~(\ref{eq:th2}) and
(\ref{eq:Tsat}) can be applied is determined by the condition that the
parallel thermal speed $v_{\parallel \rm i}$ remain~$\ll v_{\rm A}$ so
that ion Landau damping and transit-time damping of KAWs can be
neglected. Assuming that $n_0$ and $B_0$ are given by
equations~(\ref{eq:n}) and (\ref{eq:B0}), that $T_{\parallel \rm
  i}(r)$ for protons is no greater than the $r$-dependent (isotropic)
temperature in equation~(47) of \cite{cranmer05}, and that protons are
the ion species with the largest value of~$v_{\parallel \rm i}$, one
finds that $v_{\parallel \rm i} < v_{\rm A}/3$ for all ion species at
to at least~$15 R_{\sun}$.

Temperature profiles from equations~(\ref{eq:th2}) and~(\ref{eq:Tsat})
are plotted in figure~\ref{fig:Tcomp}.  The left panel of this figure
shows $T_{\perp \rm i}$ for protons and ${\rm O}^{+5}$ ions at $1.8
R_{\sun}< r < 3.3 R_{\sun}$, and the right panel shows the $T_{\perp
  \rm i}$ profiles for four ion species out to~$15 R_{\sun}$.  The
data points in both panels represent the ion kinetic temperatures
obtained by \cite{esser99} from UVCS observations. The model
temperature profiles in this figure were constructed using the $n$,
$B_0$, and $U$ profiles in equations~(\ref{eq:n})
through~(\ref{eq:defU}) and the values of $L_\perp$ and $\delta v_0$
given in equations~(\ref{eq:defLperp}) and (\ref{eq:dv0b}) with
$\delta v_{\rm a} = 155$~km/s as in figure~\ref{fig:swv_corona}.  The
value $c_2 = 0.15$ was chosen to match the observed ${\rm O}^{+5}$
temperatures at $ r> 1.8 R_{\sun}$.  The value $\alpha_{\rm p} = 0.71$
was then chosen to match the observed proton temperatures at $ r>1.8
R_{\sun}$.

\begin{figure*}[t]
\centerline{
\includegraphics[width=8.cm]{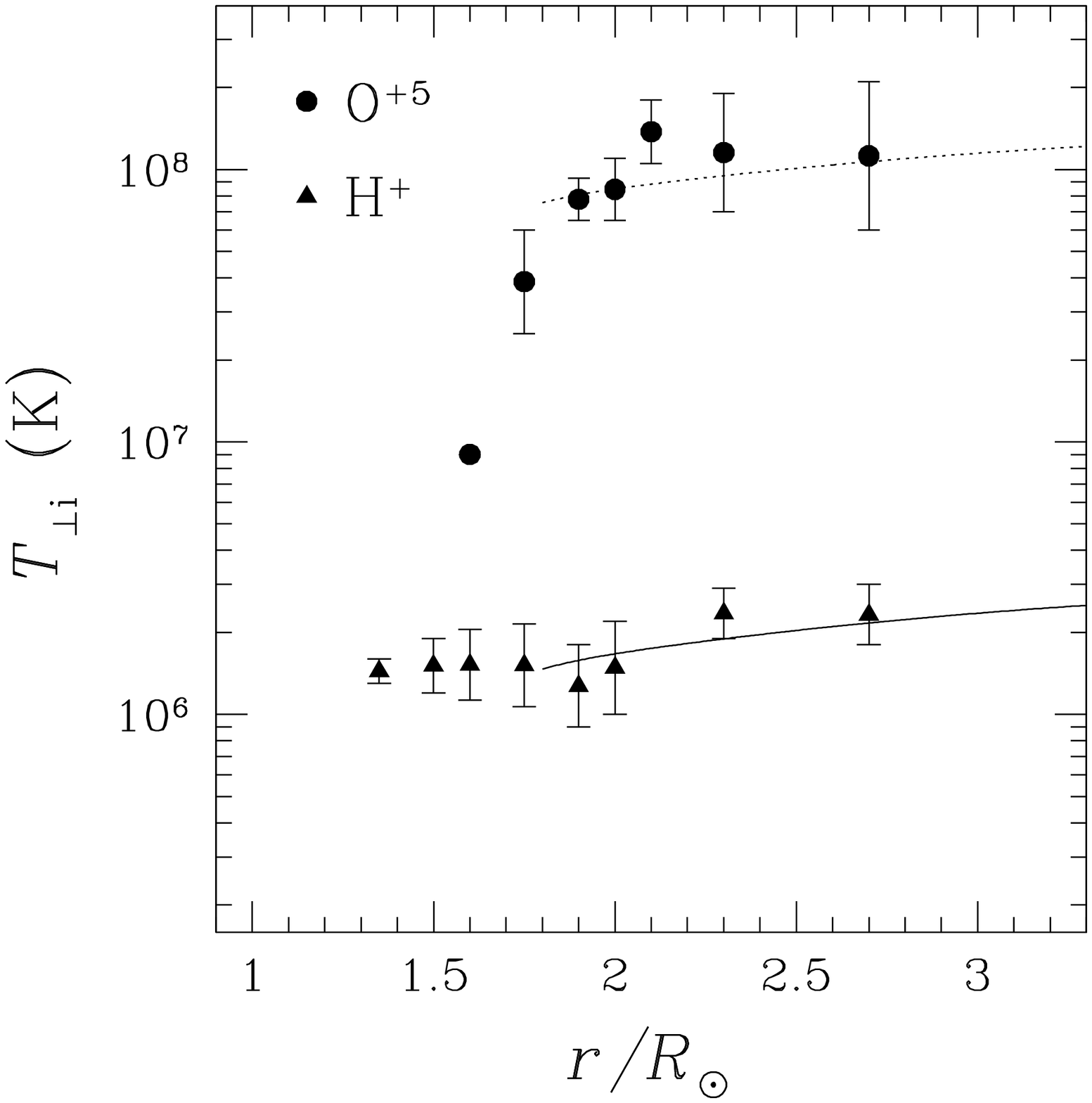}
\hspace{0.5cm} 
\includegraphics[width=8.cm]{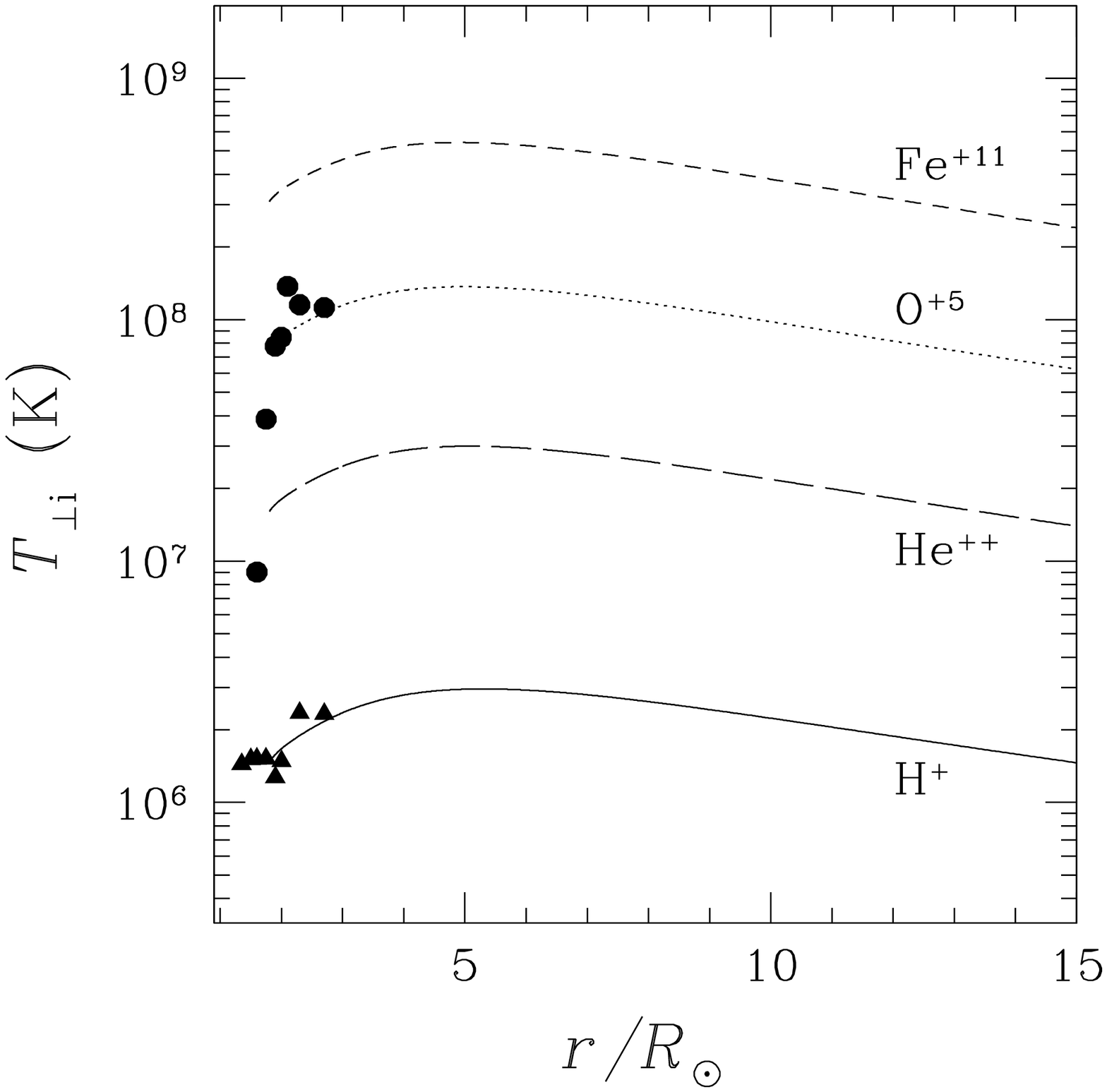}
}
\caption{{\em Left panel:}
the solid (dotted) line is $T_{\perp \rm i}$ for protons (${\rm O}^{+5}$) from 
equations~(\ref{eq:th2}) and (\ref{eq:Tsat}).
{\em Right panel:} the different curves show
$T_{\perp \rm i}$  from equations~(\ref{eq:th2}) and (\ref{eq:Tsat})
for four ion species over a larger range in~$r$.
{\em Both panels:} equations~(\ref{eq:th2}) and (\ref{eq:Tsat})
are evaluated with $\alpha_{\rm i} =
  0.71$ for protons and~$\alpha_{\rm i} = 1$ for other ions.  The
  value of $U_{\rm i}$ is set equal to~$U$ in equation~(\ref{eq:defU})
  for protons and $1.75U$ for other ions.  For all curves, $c_{2} =
  0.15$ and $a = 0.25$, $L_\perp $ and $\delta v_0$ are taken from
  equations~(\ref{eq:defLperp}) and (\ref{eq:dv0b}) with $\delta
  v_{\rm a} = 155$~km/s, and $n$ and $B$ are taken from
  equations~(\ref{eq:n}) and~(\ref{eq:B0}).  The circles and triangles
  correspond to observed kinetic temperatures for ${\rm O}^{+5}$ ions and
  protons from \cite{esser99}, with the error bars included in the left panel only.
  \label{fig:Tcomp}}\vspace{0.5cm}
\end{figure*}

Possible sources of error in figure~\ref{fig:Tcomp} include
uncertainties in the values of~$c_2$, $a$, $\alpha_{\rm i}$,
and~$U_{\rm i}(r)$.  For example, if $c_2$ is increased from~0.15
to~0.34, but all other model parameters are held fixed, then $T_{\perp
  \rm p}$ decreases by a factor of~4.6 at~$r=2R_{\sun}$.  If $a$ is
increased from~1/4 to~1/3 but all other parameters are fixed at their
original values, then $T_{\perp \rm p}$ decreases by a factor of~27
at~$r=2R_{\sun}$.  If $\alpha_{\rm i}$ is increased from~0.71 to~1 for
protons, then $T_{\perp \rm p}$ increases by a factor of~2.5
at~$r=2R_{\sun}$. The value of~$U_{\rm i}(r)$ has less of an impact on
the ion temperatures that follow from equations~(\ref{eq:th2}) and
(\ref{eq:Tsat}).  If $U_{\rm i}$ is decreased from~$1.75U$ to~$1.5U$
for~${\rm O}^{+5}$ ions, then $T_{\perp \rm i}$ increases by only
5.2\% at $r=2R_{\sun}$.

There is, however, a potentially larger source of error associated
with~$U_{\rm i}(r)$.  As the relative velocity $\Delta U_{\rm i}$
between heavy ions and protons approaches~$v_{\rm A}$, the electric
field in the heavy-ion frame decreases, since most of
the Alfv\'en-wave fluctuations propagate away from the Sun in the
proton frame. As a result, the stochastic heating rate~$Q_{\rm i}$ for
heavy ions decreases~\citep{chandran10}.  The acceleration of heavy
ions to relative flow speeds $\sim v_{\rm A}$ provides a second
possible mechanism (in addition to the reduction of~$\epsilon_{\rm i}$
through the increase of~$T_{\perp \rm i}$) for saturating stochastic
heating, provided the amplitudes of sunward-propagating waves are much
less than the amplitudes of anti-Sunward waves at $\lambda_\perp \sim
\rho_{\rm i}$.  The effects of~$\Delta U_{\rm i}$ on~$Q_{\perp \rm i}$
are neglected in figure~\ref{fig:Tcomp}, which is justified at
$r\lesssim 3 R_{\sun}$, where UVCS observations show that~$\Delta
U_{\rm i} \ll v_{\rm A}$~\citep{kohl98}.  However, for the velocity
profiles assumed in the construction of figure~\ref{fig:Tcomp},
$\Delta U_{\rm i} \sim v_{\rm A}$ near~$r= 15 R_{\sun}$, suggesting
that the heavy ion temperatures near~$r= 15 R_{\sun}$ may be
over-estimated by equation~(\ref{eq:Tsat}).

\begin{figure}[t]
\centerline{\includegraphics[width=8.cm]{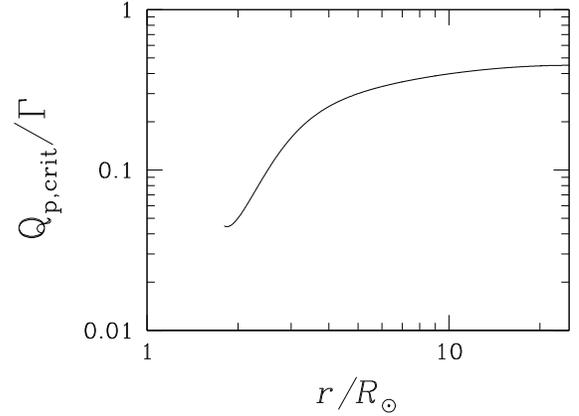}}
\caption{$Q_{\rm p, crit}$ is the approximate heating rate required to sustain protons at the temperature in the right panel of figure~\ref{fig:Tcomp} given the outflow speed in equation~(\ref{eq:defU}). $\Gamma$ is the turbulent dissipation rate in equation~(\ref{eq:Gamma0}).
  \label{fig:HF}}\vspace{0.5cm}
\end{figure}

A self-consistency check on the assumed value of~$\alpha_{\rm i}$ for
protons can be obtained by comparing the cascade power~$\Gamma$ with
the heating rate $Q_{\rm p, crit} = k_{\rm B} T_{\perp \rm p} /(m_{\rm
  p} t_{\rm exp})$ required to sustain protons at the 
temperature~$T_{\perp \rm p}$ given by equations~(\ref{eq:th2}) and
(\ref{eq:Tsat}). If the condition $Q_{\rm p, crit} < \Gamma$ were not
satisfied, then protons would absorb so much energy from the cascade
that $\alpha_{\rm p}$ would drop below the assumed value of~0.71,
thereby reducing~$Q_{\rm p, crit}$ below~$\Gamma$ as required by
energy conservation. The value of $Q_{\rm p, crit}/\Gamma$ is shown in
figure~\ref{fig:HF} for $ 1.8 R_{\sun}< r < 15 R_{\sun}$ for the same
set of assumptions used to obtain the proton temperature profile in
figure~\ref{fig:Tcomp}. It can be seen that $Q_{\rm p, crit} < \Gamma$
throughout this range of~$r$. However, some caution is warranted
here. As noted by \cite{chandran09c}, equation~(\ref{eq:Gamma0})
likely overestimates the turbulent dissipation rate for two
reasons. First, large-scale Alfv\'en waves launched from the Sun must
propagate some distance into the corona before their energy cascades
all the way to the dissipation scale. Second,
equation~(\ref{eq:Gamma0}) is derived in the limit of
small~$L_\perp$. As shown in figure~3 of~\cite{chandran09c} (and in
figure~5 of \cite{dmitruk02} for the case~$U=0$), finite values
of~$L_\perp$ reduce~$\Gamma$ relative to the small-$L_\perp$ limit.
These effects are particularly important close to the Sun.  The small
value of $Q_{\rm p, crit} /\Gamma$ at $r\sim 2 R_{\sun}$ shown in
figure~\ref{fig:HF} may thus significantly underestimate the fraction
of the cascade power absorbed by protons at this location.

\vspace{0.2cm} 
\section{Conclusion}
\label{sec:conclusion} 
\vspace{0.2cm} 

Low-frequency Alfv\'en-wave/KAW turbulence offers a promising
explanation for the detailed features of ion kinetic temperatures seen
in UVCS observations of coronal holes, including the abrupt radial
variations in the ${\rm O}^{+5}$ temperature profile, the widely
different temperatures of protons and ${\rm O}^{+5}$ ions, and the
large ${\rm O}^{+5}$ temperature anisotropy ($T_{\perp \rm i} \gg
T_{\parallel \rm i}$). When $\beta \ll 1$, ion heating from the linear
damping of low-frequency Alfv\'en waves and KAWs can be neglected, and
stochastic heating is arguably the primary way in which low-frequency
Alfv\'en-wave/KAW turbulence heats ions.  As shown
by~\cite{chandran10}, the stochastic heating rate is a strongly
decreasing function of~$T_{\perp \rm i}$. At small $v_{\perp \rm i}$,
$\epsilon_{\rm i}$~is comparatively large and ion gyro orbits are
strongly perturbed as the ions traverse the electrostatic potential of
the gyro-scale fluctuations. This leads to chaotic ion orbits and
strong stochastic heating.  In contrast, at large~$v_{\perp \rm i}$,
ion gyromotion is only weakly perturbed by the electrostatic potential
of the turbulent fluctuations, ions drift smoothly with nearly
circular orbits in the plane perpendicular to~${\bf B}$, and
stochastic heating is weak.  As discussed in section~\ref{sec:corona},
the observed ${\rm O}^{+5}$ temperature increases rapidly from $\sim
10^7$~K to $\sim 10^8$~K between $r=1.6 R_{\sun}$ and $r=1.9
R_{\sun}$, as the ions leave the region at $r\lesssim 1.6R_{\sun}$ in
which collisional energy exchange with protons is
important~\citep{esser99}. At $r= 1.9 R_{\sun}$, the ${\rm O}^{+5}$
temperature profile abruptly flattens, and $T_{\perp \rm i}$
remains~$\sim 10^8$~K out to~$2.7 R_{\sun}$, the largest radius
observed.  This temperature profile is consistent with rapid
stochastic heating of ${\rm O}^{+5}$ at $T_{\perp \rm i} \sim 10^7$~K
that saturates at $T_{\perp \rm i} \sim 10^8$~K due to the reduction
in orbit stochasticity.
 
Ion thermal conduction, like collisional energy exchange between
particle species, can be neglected to a good approximation
at~$r\gtrsim 2 R_{\sun}$. As a result, the strong temperature
dependence of the heating rate causes all minor ion temperatures to
evolve at $r\gtrsim 2 R_{\sun}$ to values at which $t_{\rm h} \sim
t_{\rm exp}$; at lower~$T_{\perp \rm i}$, minor ions are rapidly
heated, while at higher temperatures heating is ineffective and ions
cool adiabatically.  Because of the exponential dependence of $t_{\rm
  h}$ on $\epsilon_{\rm i}$, the condition $t_{\rm h} \sim t_{\rm
  exp}$ is satisfied by different ion species at nearly equal values
of~$\epsilon_{\rm i}$.  This then leads to slightly more than
mass-proportional temperatures --- in particular, $T_{\perp \rm i}
\propto A\cdot(A/Z)^l$ for ions with the same values of $\epsilon_{\rm
  i}$ and $\alpha_{\rm i}$, as in equations~(\ref{eq:Tsat}) and
(\ref{eq:Tscaling}).  The proton temperature is even smaller relative
to minor-ion temperatures than the scaling in equation~(\ref{eq:Tsat})
with constant~$\alpha_{\rm i}$ would suggest, because proton heating
and electron Landau damping reduce the amplitude of the fluctuations
at $\lambda_\perp \sim \rho_{\rm p}$, which make the largest
contribution to the stochastic proton heating rate. The condition
$t_{\rm h} \sim t_{\rm exp}$ thus leads to ${\rm O}^{+5}$ temperatures
that are much higher than the proton temperature, consistent with UVCS
observations. As described briefly in section~\ref{sec:heating},
stochastic heating primarily increases the speed of ion thermal
motions perpendicular to~${\bf B}$ when~$\beta \ll
1$~\citep{chandran10}. Because of this, stochastic heating also offers
an explanation for the observation that $T_{\perp \rm i} \gg
T_{\parallel \rm i}$ for~${\rm
  O}^{+5}$~ions~\citep{kohl98,li98,antonucci00}.

In section~\ref{sec:corona}, $T_{\perp \rm i}$ profiles are calculated
for several ion species from the condition $t_{\rm h} \sim t_{\rm
  exp}$ using an observationally constrained model of Alfv\'en-wave
turbulence in coronal holes. For plausible values of the model
parameters, the resulting temperature profiles provide a good match to
observations of protons and ${\rm O}^{+5}$ ions.  However, there are
several sources of uncertainty in these calculations, the largest of
which are associated with the parameters~$c_2$, $a$, and~$\alpha_{\rm
  p}$ (the value of~$\alpha_{\rm i}$ for protons). The constant~$c_2$
relates to the efficiency of stochastic ion heating and was evaluated
by \cite{chandran10} using numerical simulations of test-particle
protons interacting with randomly phased waves in a low-$\beta$
plasma.  However, the value of~$c_2$ for strong turbulence and for
other ion species is not yet known. The quantities~$a$
and~$\alpha_{\rm p}$ describe, respectively, the slope of the
inertial-range velocity power spectrum and the decrement in the
velocity spectrum at~$\lambda_\perp \sim \rho_{\rm p}$ arising from
electron and ion damping.  In order to determine whether stochastic
heating can indeed explain the ion temperatures observed in coronal
holes, future work is needed to determine $c_2$, $a$, and~$\alpha_{\rm
  p}$ more accurately. For example, direct numerical simulations or a
cascade model (see, e.g., \cite{howes08a}) of strong AW/KAW turbulence
including both stochastic heating and electron Landau damping would
help ascertain the value of~$\alpha_{\rm p}$. Direct numerical
simulations of strong, highly anisotropic, AW/KAW turbulence
interacting with test particles could be used to determine the value
of~$c_2$ that is appropriate for coronal holes.  In addition, {\em in
  situ} measurements from NASA's planned {\em Solar Probe} mission
will establish the power spectrum of AW/KAW turbulence and determine
$U_{\rm i}(r)$ and $T_{\perp \rm i}(r)$ for several ion species at
heliocentric distances as small as~$\sim 9.5R_{\sun}$.  These
measurements will lead to a rigorous test of the stochastic heating
model, including the $T_{\perp \rm i}(r)$ predictions shown in
figure~\ref{fig:Tcomp} and the dependence of~$T_{\perp \rm i}$ on ion
mass and charge given in equation~(\ref{eq:Tscaling}).

\acknowledgements I thank Joe Hollweg, Phil Isenberg, Eliot Quataert,
and the referee for valuable suggestions and feedback on an earlier
version of this manuscript, and Peter Bochsler, Bo Li, and Barrett
Rogers for helpful discussions. This work was supported in part by NSF
Grants AST-0613622, AGS-0851005, and AGS-1003451, DOE Grant
DE-FG02-07-ER46372, and NASA Grants NNX07AP65G and NNX08AH52G.

\bibliography{articles}

\begin{thebibliography}{74}
\expandafter\ifx\csname natexlab\endcsname\relax\def\natexlab#1{#1}\fi

\bibitem[{{Antonucci} {et~al.}(2000){Antonucci}, {Dodero}, \&
  {Giordano}}]{antonucci00}
{Antonucci}, E., {Dodero}, M.~A., \& {Giordano}, S. 2000, \solphys, 197, 115

\bibitem[{{Bale} {et~al.}(2005){Bale}, {Kellogg}, {Mozer}, {Horbury}, \&
  {Reme}}]{bale05}
{Bale}, S.~D., {Kellogg}, P.~J., {Mozer}, F.~S., {Horbury}, T.~S., \& {Reme},
  H. 2005, Physical Review Letters, 94, 215002

\bibitem[{{Bame} {et~al.}(1977){Bame}, {Asbridge}, {Feldman}, \&
  {Gosling}}]{bame77}
{Bame}, S.~J., {Asbridge}, J.~R., {Feldman}, W.~C., \& {Gosling}, J.~T. 1977,
  \jgr, 82, 1487

\bibitem[{{Belcher} \& {Davis}(1971)}]{belcher71}
{Belcher}, J.~W., \& {Davis}, Jr., L. 1971, \jgr, 76, 3534

\bibitem[{{Boldyrev}(2006)}]{boldyrev06}
{Boldyrev}, S. 2006, Physical Review Letters, 96, 115002

\bibitem[{{Bourouaine} {et~al.}(2008){Bourouaine}, {Marsch}, \&
  {Vocks}}]{bourouaine08}
{Bourouaine}, S., {Marsch}, E., \& {Vocks}, C. 2008, \apjl, 684, L119

\bibitem[{{Bruno} \& {Carbone}(2005)}]{bruno05}
{Bruno}, R., \& {Carbone}, V. 2005, Living Reviews in Solar Physics, 2, 4

\bibitem[{{Chandran} \& {Hollweg}(2009)}]{chandran09c}
{Chandran}, B.~D.~G., \& {Hollweg}, J.~V. 2009, \apj, 707, 1659

\bibitem[{{Chandran} {et~al.}(2010){Chandran}, {Li}, {Rogers}, {Quataert}, \&
  {Germaschewski}}]{chandran10}
{Chandran}, B.~D.~G., {Li}, B., {Rogers}, B.~N., {Quataert}, E., \&
  {Germaschewski}, K. 2010, ApJ, submitted, (arXiv:1001.2069)

\bibitem[{{Chandran} {et~al.}(2009){Chandran}, {Quataert}, {Howes}, {Xia}, \&
  {Pongkitiwanichakul}}]{chandran09d}
{Chandran}, B.~D.~G., {Quataert}, E., {Howes}, G.~G., {Xia}, Q., \&
  {Pongkitiwanichakul}, P. 2009, \apj, 707, 1668

\bibitem[{{Chen} {et~al.}(2001){Chen}, {Lin}, \& {White}}]{chen01}
{Chen}, L., {Lin}, Z., \& {White}, R. 2001, Physics of Plasmas, 8, 4713

\bibitem[{{Cho} {et~al.}(2002){Cho}, {Lazarian}, \& {Vishniac}}]{cho02b}
{Cho}, J., {Lazarian}, A., \& {Vishniac}, E.~T. 2002, \apj, 564, 291

\bibitem[{{Cho} \& {Vishniac}(2000)}]{cho00}
{Cho}, J., \& {Vishniac}, E.~T. 2000, \apj, 539, 273

\bibitem[{{Coleman}(1968)}]{coleman68}
{Coleman}, P.~J. 1968, \apj, 153, 371

\bibitem[{{Cranmer}(2010)}]{cranmer10}
{Cranmer}, S.~R. 2010, \apj, 710, 676

\bibitem[{{Cranmer} {et~al.}(1999){Cranmer}, {Field}, \& {Kohl}}]{cranmer99}
{Cranmer}, S.~R., {Field}, G.~B., \& {Kohl}, J.~L. 1999, \apj, 518, 937

\bibitem[{{Cranmer} {et~al.}(2009){Cranmer}, {Matthaeus}, {Breech}, \&
  {Kasper}}]{cranmer09}
{Cranmer}, S.~R., {Matthaeus}, W.~H., {Breech}, B.~A., \& {Kasper}, J.~C. 2009,
  \apj, 702, 1604

\bibitem[{{Cranmer} \& {van Ballegooijen}(2003)}]{cranmer03}
{Cranmer}, S.~R., \& {van Ballegooijen}, A.~A. 2003, \apj, 594, 573

\bibitem[{{Cranmer} \& {van Ballegooijen}(2005)}]{cranmer05}
---. 2005, 156, 265

\bibitem[{{Cranmer} {et~al.}(2007){Cranmer}, {van Ballegooijen}, \&
  {Edgar}}]{cranmer07}
{Cranmer}, S.~R., {van Ballegooijen}, A.~A., \& {Edgar}, R.~J. 2007, \apjs,
  171, 520

\bibitem[{{De Pontieu} {et~al.}(2007){De Pontieu}, {McIntosh}, {Carlsson},
  {Hansteen}, {Tarbell}, {Schrijver}, {Title}, {Shine}, {Tsuneta}, {Katsukawa},
  {Ichimoto}, {Suematsu}, {Shimizu}, \& {Nagata}}]{depontieu07}
{De Pontieu}, B., {et~al.} 2007, Science, 318, 1574

\bibitem[{{Dmitruk} {et~al.}(2002){Dmitruk}, {Matthaeus}, {Milano}, {Oughton},
  {Zank}, \& {Mullan}}]{dmitruk02}
{Dmitruk}, P., {Matthaeus}, W.~H., {Milano}, L.~J., {Oughton}, S., {Zank},
  G.~P., \& {Mullan}, D.~J. 2002, \apj, 575, 571

\bibitem[{{Dmitruk} {et~al.}(2004){Dmitruk}, {Matthaeus}, \&
  {Seenu}}]{dmitruk04}
{Dmitruk}, P., {Matthaeus}, W.~H., \& {Seenu}, N. 2004, \apj, 617, 667

\bibitem[{{Drake} {et~al.}(2009){Drake}, {Cassak}, {Shay}, {Swisdak}, \&
  {Quataert}}]{drake09}
{Drake}, J.~F., {Cassak}, P.~A., {Shay}, M.~A., {Swisdak}, M., \& {Quataert},
  E. 2009, \apjl, 700, L16

\bibitem[{{Esser} {et~al.}(1999){Esser}, {Fineschi}, {Dobrzycka}, {Habbal},
  {Edgar}, {Raymond}, {Kohl}, \& {Guhathakurta}}]{esser99}
{Esser}, R., {Fineschi}, S., {Dobrzycka}, D., {Habbal}, S.~R., {Edgar}, R.~J.,
  {Raymond}, J.~C., {Kohl}, J.~L., \& {Guhathakurta}, M. 1999, \apjl, 510, L63

\bibitem[{{Feldman} {et~al.}(1997){Feldman}, {Habbal}, {Hoogeveen}, \&
  {Wang}}]{feldman97}
{Feldman}, W.~C., {Habbal}, S.~R., {Hoogeveen}, G., \& {Wang}, Y. 1997, \jgr,
  102, 26905

\bibitem[{{Galtier} {et~al.}(2000){Galtier}, {Nazarenko}, {Newell}, \&
  {Pouquet}}]{galtier00}
{Galtier}, S., {Nazarenko}, S.~V., {Newell}, A.~C., \& {Pouquet}, A. 2000,
  Journal of Plasma Physics, 63, 447

\bibitem[{{Goldreich} \& {Sridhar}(1995)}]{goldreich95}
{Goldreich}, P., \& {Sridhar}, S. 1995, \apj, 438, 763

\bibitem[{{Goldstein} {et~al.}(1995){Goldstein}, {Roberts}, \&
  {Matthaeus}}]{goldstein95a}
{Goldstein}, M.~L., {Roberts}, D.~A., \& {Matthaeus}, W.~H. 1995, \araa, 33,
  283

\bibitem[{{Grappin} {et~al.}(1990){Grappin}, {Mangeney}, \&
  {Marsch}}]{grappin90}
{Grappin}, R., {Mangeney}, A., \& {Marsch}, E. 1990, \jgr, 95, 8197

\bibitem[{{Harmon} \& {Coles}(2005)}]{harmon05}
{Harmon}, J.~K., \& {Coles}, W.~A. 2005, Journal of Geophysical Research (Space
  Physics), 110, 3101

\bibitem[{{Haugen} {et~al.}(2004){Haugen}, {Brandenburg}, \&
  {Dobler}}]{haugen04}
{Haugen}, N.~E., {Brandenburg}, A., \& {Dobler}, W. 2004, \pre, 70, 016308

\bibitem[{{Heinemann} \& {Olbert}(1980)}]{heinemann80}
{Heinemann}, M., \& {Olbert}, S. 1980, \jgr, 85, 1311

\bibitem[{{Hollweg}(1999)}]{hollweg99c}
{Hollweg}, J.~V. 1999, \jgr, 104, 14811

\bibitem[{{Hollweg} \& {Isenberg}(2007)}]{hollweg07}
{Hollweg}, J.~V., \& {Isenberg}, P.~A. 2007, Journal of Geophysical Research
  (Space Physics), 112, 8102

\bibitem[{{Howes} {et~al.}(2008a){Howes}, {Cowley}, {Dorland}, {Hammett},
  {Quataert}, \& {Schekochihin}}]{howes08a}
{Howes}, G.~G., {Cowley}, S.~C., {Dorland}, W., {Hammett}, G.~W., {Quataert},
  E., \& {Schekochihin}, A.~A. 2008a, Journal of Geophysical Research (Space
  Physics), 113, 5103

\bibitem[{{Howes} {et~al.}(2008b){Howes}, {Dorland}, {Cowley}, {Hammett},
  {Quataert}, {Schekochihin}, \& {Tatsuno}}]{howes08b}
{Howes}, G.~G., {Dorland}, W., {Cowley}, S.~C., {Hammett}, G.~W., {Quataert},
  E., {Schekochihin}, A.~A., \& {Tatsuno}, T. 2008b, Physical Review Letters,
  100, 065004

\bibitem[{{Isenberg} \& {Vasquez}(2009)}]{isenberg09}
{Isenberg}, P.~A., \& {Vasquez}, B.~J. 2009, \apj, 696, 591

\bibitem[{{Johnson} \& {Cheng}(2001)}]{johnson01}
{Johnson}, J.~R., \& {Cheng}, C.~Z. 2001, \grl, 28, 4421

\bibitem[{{Kasper} {et~al.}(2007){Kasper}, {Stevens}, {Lazarus}, {Steinberg},
  \& {Ogilvie}}]{kasper07}
{Kasper}, J.~C., {Stevens}, M.~L., {Lazarus}, A.~J., {Steinberg}, J.~T., \&
  {Ogilvie}, K.~W. 2007, \apj, 660, 901

\bibitem[{{Kohl, J., et al.}(1998)}]{kohl98}
{Kohl, J., et al.} 1998, \apjl, 501, L127

\bibitem[{{Kruskal}(1962)}]{kruskal62}
{Kruskal}, M. 1962, Journal of Mathematical Physics, 3, 806

\bibitem[{{Leamon} {et~al.}(1999){Leamon}, {Smith}, {Ness}, \&
  {Wong}}]{leamon99}
{Leamon}, R.~J., {Smith}, C.~W., {Ness}, N.~F., \& {Wong}, H.~K. 1999, \jgr,
  104, 22331

\bibitem[{{Lehe} {et~al.}(2009){Lehe}, {Parrish}, \& {Quataert}}]{lehe09}
{Lehe}, R., {Parrish}, I.~J., \& {Quataert}, E. 2009, \apj, 707, 404

\bibitem[{{Li} {et~al.}(1998){Li}, {Habbal}, {Kohl}, \& {Noci}}]{li98}
{Li}, X., {Habbal}, S.~R., {Kohl}, J., \& {Noci}, G. 1998, \apjl, 501, L133+

\bibitem[{{Markovskii} {et~al.}(2006){Markovskii}, {Vasquez}, {Smith}, \&
  {Hollweg}}]{markovskii06}
{Markovskii}, S.~A., {Vasquez}, B.~J., {Smith}, C.~W., \& {Hollweg}, J.~V.
  2006, \apj, 639, 1177

\bibitem[{{Maron} \& {Goldreich}(2001)}]{maron01}
{Maron}, J., \& {Goldreich}, P. 2001, \apj, 554, 1175

\bibitem[{{Marsch} {et~al.}(2004){Marsch}, {Ao}, \& {Tu}}]{marsch04}
{Marsch}, E., {Ao}, X.-Z., \& {Tu}, C.-Y. 2004, Journal of Geophysical Research
  (Space Physics), 109, 4102

\bibitem[{{Marsch} {et~al.}(1982){Marsch}, {Rosenbauer}, {Schwenn},
  {Muehlhaeuser}, \& {Neubauer}}]{marsch82a}
{Marsch}, E., {Rosenbauer}, H., {Schwenn}, R., {Muehlhaeuser}, K., \&
  {Neubauer}, F.~M. 1982, \jgr, 87, 35

\bibitem[{{Mason} {et~al.}(2008){Mason}, {Cattaneo}, \& {Boldyrev}}]{mason08}
{Mason}, J., {Cattaneo}, F., \& {Boldyrev}, S. 2008, \pre, 77, 036403

\bibitem[{{Matthaeus} {et~al.}(1999){Matthaeus}, {Zank}, {Oughton}, {Mullan},
  \& {Dmitruk}}]{matthaeus99b}
{Matthaeus}, W.~H., {Zank}, G.~P., {Oughton}, S., {Mullan}, D.~J., \&
  {Dmitruk}, P. 1999, \apjl, 523, L93

\bibitem[{{M{\"u}ller} \& {Grappin}(2005)}]{muller05}
{M{\"u}ller}, W., \& {Grappin}, R. 2005, Physical Review Letters, 95, 114502

\bibitem[{{M{\"u}ller} \& {Biskamp}(2000)}]{muller00}
{M{\"u}ller}, W.-C., \& {Biskamp}, D. 2000, Physical Review Letters, 84, 475

\bibitem[{{Ng} \& {Bhattacharjee}(1996)}]{ng96}
{Ng}, C.~S., \& {Bhattacharjee}, A. 1996, \apj, 465, 845

\bibitem[{{Parashar} {et~al.}(2009){Parashar}, {Shay}, {Cassak}, \&
  {Matthaeus}}]{parashar09}
{Parashar}, T.~N., {Shay}, M.~A., {Cassak}, P.~A., \& {Matthaeus}, W.~H. 2009,
  Physics of Plasmas, 16, 032310

\bibitem[{{Perez} \& {Boldyrev}(2008)}]{perez08a}
{Perez}, J.~C., \& {Boldyrev}, S. 2008, \apjl, 672, L61

\bibitem[{{Perez} \& {Boldyrev}(2009)}]{perez09a}
---. 2009, Physical Review Letters, 102, 025003

\bibitem[{{Phillips} \& {Gosling}(1990)}]{phillips90}
{Phillips}, J.~L., \& {Gosling}, J.~T. 1990, \jgr, 95, 4217

\bibitem[{{Podesta} {et~al.}(2007){Podesta}, {Roberts}, \&
  {Goldstein}}]{podesta07}
{Podesta}, J.~J., {Roberts}, D.~A., \& {Goldstein}, M.~L. 2007, \apj, 664, 543

\bibitem[{{Quataert}(1998)}]{quataert98}
{Quataert}, E. 1998, \apj, 500, 978

\bibitem[{{Sahraoui} {et~al.}(2009){Sahraoui}, {Goldstein}, {Robert}, \&
  {Khotyaintsev}}]{sahraoui09}
{Sahraoui}, F., {Goldstein}, M.~L., {Robert}, P., \& {Khotyaintsev}, Y.~V.
  2009, Physical Review Letters, 102, 231102

\bibitem[{{Schekochihin} {et~al.}(2009){Schekochihin}, {Cowley}, {Dorland},
  {Hammett}, {Howes}, {Quataert}, \& {Tatsuno}}]{schekochihin09}
{Schekochihin}, A.~A., {Cowley}, S.~C., {Dorland}, W., {Hammett}, G.~W.,
  {Howes}, G.~G., {Quataert}, E., \& {Tatsuno}, T. 2009, \apjs, 182, 310

\bibitem[{Shebalin {et~al.}(1983)Shebalin, Matthaeus, \&
  Montgomery}]{shebalin83}
Shebalin, J.~V., Matthaeus, W., \& Montgomery, D. 1983, Journal of Plasma
  Physics, 29, 525

\bibitem[{{Smith} {et~al.}(2001){Smith}, {Matthaeus}, {Zank}, {Ness},
  {Oughton}, \& {Richardson}}]{smith01}
{Smith}, C.~W., {Matthaeus}, W.~H., {Zank}, G.~P., {Ness}, N.~F., {Oughton},
  S., \& {Richardson}, J.~D. 2001, \jgr, 106, 8253

\bibitem[{{Spruit}(1981)}]{spruit81}
{Spruit}, H.~C. 1981, NASA Special Publication, 450, 385

\bibitem[{{Stawarz} {et~al.}(2009){Stawarz}, {Smith}, {Vasquez}, {Forman}, \&
  {MacBride}}]{stawarz09}
{Stawarz}, J.~E., {Smith}, C.~W., {Vasquez}, B.~J., {Forman}, M.~A., \&
  {MacBride}, B.~T. 2009, \apj, 697, 1119

\bibitem[{{Tomczyk} {et~al.}(2007){Tomczyk}, {McIntosh}, {Keil}, {Judge},
  {Schad}, {Seeley}, \& {Edmondson}}]{tomczyk07}
{Tomczyk}, S., {McIntosh}, S.~W., {Keil}, S.~L., {Judge}, P.~G., {Schad}, T.,
  {Seeley}, D.~H., \& {Edmondson}, J. 2007, Science, 317, 1192

\bibitem[{{Tu} \& {Marsch}(1995)}]{tumarsch95}
{Tu}, C., \& {Marsch}, E. 1995, Space Science Reviews, 73, 1

\bibitem[{{Velli}(1993)}]{velli93}
{Velli}, M. 1993, \aap, 270, 304

\bibitem[{{Velli} {et~al.}(1989){Velli}, {Grappin}, \& {Mangeney}}]{velli89}
{Velli}, M., {Grappin}, R., \& {Mangeney}, A. 1989, Physical Review Letters,
  63, 1807

\bibitem[{{Verdini} \& {Velli}(2007)}]{verdini07}
{Verdini}, A., \& {Velli}, M. 2007, \apj, 662, 669

\bibitem[{{Verdini} {et~al.}(2010){Verdini}, {Velli}, {Matthaeus}, {Oughton},
  \& {Dmitruk}}]{verdini10}
{Verdini}, A., {Velli}, M., {Matthaeus}, W.~H., {Oughton}, S., \& {Dmitruk}, P.
  2010, \apjl, 708, L116

\bibitem[{{Voitenko} \& {Goossens}(2004)}]{voitenko04}
{Voitenko}, Y., \& {Goossens}, M. 2004, \apjl, 605, L149

\bibitem[{{White} {et~al.}(2002){White}, {Chen}, \& {Lin}}]{white02}
{White}, R., {Chen}, L., \& {Lin}, Z. 2002, Physics of Plasmas, 9, 1890

\end{thebibliography}

\end{document}